# COMPARATIVE ANALYSIS OF PEAK CORRELATION CHARACTERISTICS OF NON-ORTHOGONAL SPREADING CODES FOR WIRELESS SYSTEMS

Dr. Deepak Kedia

Department of Electronics & Communication Engineering,
G.J. University of Science & Technology, Hisar, Haryana, India
`kedia_02in@yahoo.com`

## ABSTRACT

*The performance of a CDMA based wireless system is largely dependent on the characteristics of pseudo-random spreading codes. The spreading codes should be carefully chosen to ensure highest possible peak value of auto-correlation function and lower correlation peaks (side-lobes) at non-zero time-shifts. Simultaneously, zero cross-correlation value at all time shifts is required in order to eliminate the effect of multiple access interference at the receiver. But no such code family exists which possess both characteristics simultaneously. That's why an exhaustive effort has been made in this paper to evaluate the peak correlation characteristics of various non-orthogonal spreading codes and suggest a suitable solution.*

## KEYWORDS

*Non-Orthogonal Codes, Auto-correlation, Cross-correlation, CDMA, Pseudo-Noise (PN)*

## 1. INTRODUCTION

For future applications like 4-G [1], [2] or next generation broadband wireless mobile systems, a CDMA system [3], [4] combined with multicarrier modulation is very effective in combating severe multipath interference and providing multiple access simultaneously. Multicarrier CDMA [5] is being proposed as a strong candidate for radio access in future broadband wireless communication systems. Thus, the performance of next generation wireless systems is also dependent on the characteristics of pseudo-random spreading codes because these systems will be based on hybrid CDMA [5] techniques. The desirable characteristics of CDMA codes [6] include (i) availability of large number of codes (ii) impulsive auto-correlation function (iii) zero cross-correlation values (iv) randomness (v) ease of generation (vi) low Peak to Average Power Ratio (PAPR) value [7] and (v) support for variable data rates [8]. Hence, the choice of spreading codes in CDMA mobile systems is mainly governed by the above mentioned desirable characteristics. In this paper, various non-orthogonal codes i.e. PN codes and Gold codes have been investigated with special emphasis on their peak correlation characteristics.

Non-orthogonal spreading codes are those which give non-zero cross-correlation values for different time shifts. It results into multiple access interference (MAI) and thus limits the maximum number of users supported by CDMA system. A pseudo-noise (PN) code is a periodic binary sequence generated using linear feedback shift register (LFSR) structure [8]. These codes are also known as Maximal length sequences. Secondly, Gold codes assume significance because of their large code family size as compared to their PN counterpart. In fact, these Gold codes are constructed using a pair of PN sequences (usually preferred pair) [9].

Ideally, the auto-correlation function (ACF) of a spreading code should be impulsive and cross-correlation function should possess zero values for all time-shifts. Impulsive ACF is required at receiver side to distinguish the desired user from other users producing MAI. Thus, spreading codes should be carefully chosen to ensure highest possible peak value of auto-correlation function and lower correlation peaks (side-lobes) at non-zero shifts. On the other hand, zero cross-correlation value at all time shifts is required in order to eliminate the effect of multiple access interference at the receiver. This is so because CCF indicates the correlation between the desired code sequence and the undesired ones at the receiver. Therefore, it is desirable to have a code dictionary consisting of spreading codes which possess both impulsive ACF and all zero CCF characteristics. But no such code family exists which possess both characteristics simultaneously. Thus, it becomes a tedious task to select a spreading sequence possessing desirable characteristics for implementation of a CDMA [10] based next generation wireless mobile communication system. That's why an exhaustive effort has been made in this paper to evaluate the peak correlation characteristics of various non-orthogonal spreading codes and suggest a suitable solution.

## 2. PEAK CORRELATION CHARACTERISTICS

### 2.1. Peak Auto-Correlation

Auto-correlation is a measure of the similarity between a code $C(t)$ and its time shifted replica [6], [8]. Mathematically, it is defined as:

$$\psi_a(\tau) = \int_{-\infty}^{\infty} c(t) * c(t-\tau) dt \quad (1)$$

Ideally, this auto-correlation function (ACF) should be impulsive i.e. peak value at zero time shift and zero values at all other time-shifts (i.e. side-lobes). This is required at the receiver side for proper synchronization and to distinguish the desired user from other users producing MAI. Thus, spreading codes should be carefully chosen to ensure highest possible peak value of auto-correlation function at zero shift and lower side-lobes peaks at non-zero shifts. The larger the gap between main ACF peak (zero shift) and side-lobes peaks, the better is the performance of CDMA mobile system.

### 2.2. Peak Cross-Correlation

Like peak auto-correlation, cross-correlation is also a crucial parameter which dictates the choice of suitable spreading codes for wireless CDMA systems. Each user in CDMA based system is being assigned a separate and unique code sequence. Cross-correlation [8] is the measure of similarity between two different code sequences $C_1(t)$ and $C_2(t)$. Mathematically, it is defined as:

$$\psi_c(\tau) = \int_{-\infty}^{\infty} c_1(t) * c_2(t-\tau) dt \quad (2)$$

Cross-correlation function (CCF) in-fact indicates the correlation between the desired code sequence and the undesired ones at the receiver. Therefore, in order to eliminate the effect of multiple access interference at the receiver, the cross-correlation value must be zero at all time shifts. The codes for which $\psi_c(\tau) = 0$ i.e. zero cross-correlation value at all time shifts, are known as orthogonal codes [5], [8]. Therefore, it is desirable to have a code dictionary consisting of spreading codes which possess both impulsive ACF and all zero CCF characteristics. But unfortunately, no such code family exists which possess both characteristics simultaneously. Therefore, a communication engineer has to compromise with maximum possible difference between ACF peak and CCF peak of the codes selected.

It is pertinent to mention here that the envelope power of a multicarrier [11] CDMA signal $S(t)$

is also significantly affected by these auto-correlation and cross-correlation functions of the selected code sequences. Mathematically, the relationship between the envelope power $|S(t)|^2$ and correlation functions [12] may be expressed as:

$$|S(t)|^2 = L + \frac{2}{N} \text{Re}\left[\sum_{n=1}^{N-1}(A[n]+X[n]).e^{j2\Pi Fn\frac{t}{T}}\right] \quad (3)$$

where $L$ is the number of simultaneously used spreading codes of $N$ chips each, $F$ is sub-carrier separation parameter and $A[n]$ & $X[n]$ are collective aperiodic auto-correlation and cross-correlation respectively as defined below.

$$A[n] \stackrel{\Delta}{=} \sum_{l=0}^{L-1} A_l[n] \quad for\ n \neq 0$$

$$X[n] \stackrel{\Delta}{=} \sum_{l=0}^{L-1} \sum_{l'=0, l' \neq l}^{L-1} b_l b_{l'}^* X_{l,l'}[n] \quad (4)$$

where $A_l[n]$ represents the aperiodic auto-correlation of $l^{th}$ spreading code, defined by the following relation:

$$A_l[n] \stackrel{\Delta}{=} \sum_{i=0}^{N-n-1} C_l[i] C_l^*[i+n] \quad (5)$$

And $X_{l,l'}[n]$ represents aperiodic cross-correlation between $l^{th}$ and $(l')^{th}$ spreading codes respectively as defined below:

$$X_{l,l'}[n] \stackrel{\Delta}{=} \sum_{i=0}^{N-n-1} C_l[i] C_{l'}^*[i+n] \quad (6)$$

It is quite evident from the Equation 3 that the envelope power and thus PAPR [13] is dependent on ACF and CCF of the chosen spreading codes for next generation multicarrier CDMA based wireless systems [14], [16], [17]. Therefore, the spreading codes which provide lower PAPR value must be selected. And, it is again the correlation characteristics which dictate the choice of spreading codes providing lower PAPR values.

## 3. GENERATION OF NON-ORTHOGONAL SPREADING CODES

### 3.1. Pseudo-Noise (PN) Codes

A pseudo-noise (PN) code is a periodic binary sequence generated using linear feedback shift register (LFSR) structure [5], [6], [8] as shown in Figure 1. These codes are also known as Maximal length sequences (m-sequences).

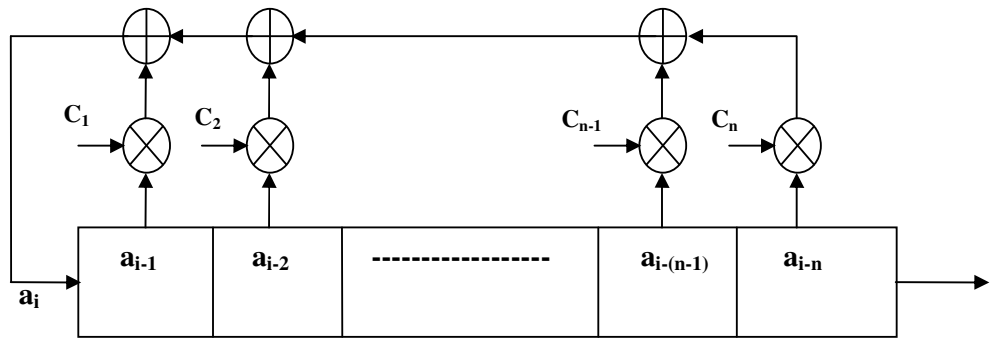

Figure 1. PN-Sequence Generator Structure

The sequence $a_i$ is generated according to the recursive formula [8]:

$$a_i = C_1 a_{i-1} + C_2 a_{i-2} + \ldots\ldots\ldots + C_n a_{i-n} = \sum_{k=1}^{n} C_k a_{i-k} \qquad (7)$$

Here, all terms are binary (0 or 1), and addition and multiplications are modulo-2. The connection vector $C_1, C_2, \ldots\ldots C_n$ defines the characteristic polynomial of the linear feedback shift register (LFSR) sequence generator and determines the main characteristics of the generated sequence. The feedback logic i.e. characteristic polynomial required for generation of different length codes has already been compiled in the literature [6]. The PN code so generated using LFSR with n number of flip-flops is periodic with period $N = 2^n - 1$.

In this paper, the above mentioned generator structure was implemented through MATLAB programming. The PN codes having length N = 7 to 255 were generated and analyzed through programming. The feedback polynomials being used for the sequence generator structure to generate the above mentioned m-sequences are listed below:

| | | |
|---|---|---|
| 7- Length PN sequence: | $X^3+X^2+1$; | |
| | $X^3+X+1$ | (Preferred Pair) |
| 15- Length PN sequence: | $X^4+X^3+1$; | |
| | $X^4+X+1$ | (Non-Preferred Pair) |
| 31- Length PN sequence: | $X^5+X^4+X^3+X^2+1$; | |
| | $X^5+X^4+X^3+X+1$ | (Preferred Pair) |
| 63- Length PN sequence: | $X^6+X+1$; | |
| | $X^6+X^5+X^2+X+1$ | (Preferred Pair) |
| 127- Length PN sequence: | $X^7+X+1$; | |
| | $X^7+X^3+1$ | (Preferred Pair) |
| 255- Length PN sequence: | $X^8+X^6+X^5+X^3+1$; | |
| | $X^8+X^4+X^3+X^2+1$ | (Non-Preferred Pair) |

Here, preferred pair of polynomials will generate PN codes having 3-valued cross-correlation function (to be discussed later). The schematic used to generate one of the above PN sequences (N=127) using characteristic polynomial $X^7+X^3+1$ is shown below in Figure 2 for illustration purpose. All the array elements a[i]'s are initialized to binary '1'. Then, with each clock cycle the feedback calculations (modulo-2 additions) are done, values in the array elements are shifted towards right and one element of the code sequence is obtained.

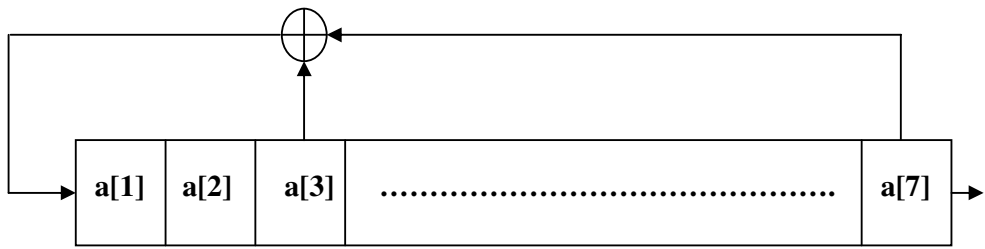

Figure 2. 127 Length PN Code ($X^7+X^3+1$)

It is thus observed that the generation structure of PN codes is very simple but family size of PN codes of different lengths is very small. As already mentioned, PN codes of each length are defined by their corresponding characteristic feedback polynomials. The PN code family size for lengths N = 7 to 1023 are tabulated in Table 1. Besides this, computer simulation was done to identify and verify all the characteristic polynomials for each of these lengths. Thus, the biggest disadvantage of PN codes is the small code family size and as a result lesser number of mobile users is supported [15].

Table 1. Possible Number of PN Codes for various Lengths

| PN Code Length | Code Family Size |
| --- | --- |
| 7 | 02 |
| 15 | 02 |
| 31 | 06 |
| 63 | 06 |
| 127 | 18 |
| 255 | 16 |
| 511 | 48 |
| 1023 | 60 |

### 3.2. Gold Codes

Gold codes assume significance because of their large code family size as compared to their PN counterpart. In fact, these Gold codes are constructed using a pair of PN sequences (usually preferred pair) [6], [8], [9]. The PN codes designed in the previous subsection are used to construct the Gold codes of desired length. Let a and $a^1$ represent a preferred pair of PN sequences having period $N= 2^n-1$. In order to generate a set of all possible Gold codes for a given length; one of the above two PN codes is delayed by one chip at a time to generate a new Gold code. Thus, the family of Gold codes is defined by {a, $a^1$, a+$a^1$, a+D$a^1$, a+$D^2a^1$,..., a+$D^{N-1}a^1$ }, where D is the delay element. With the exception of sequences 'a' and '$a^1$', the set of Gold sequences are not maximal sequences. Thus, N number of Gold codes can be generated from a preferred pair of PN sequences of length N. Hence, by including this pair of generating PN codes in the Gold code family, the total number of Gold codes becomes N+2 for each length. An illustration of how a Gold code ($2^7$-1 length) is being constructed is shown in Figure 3.

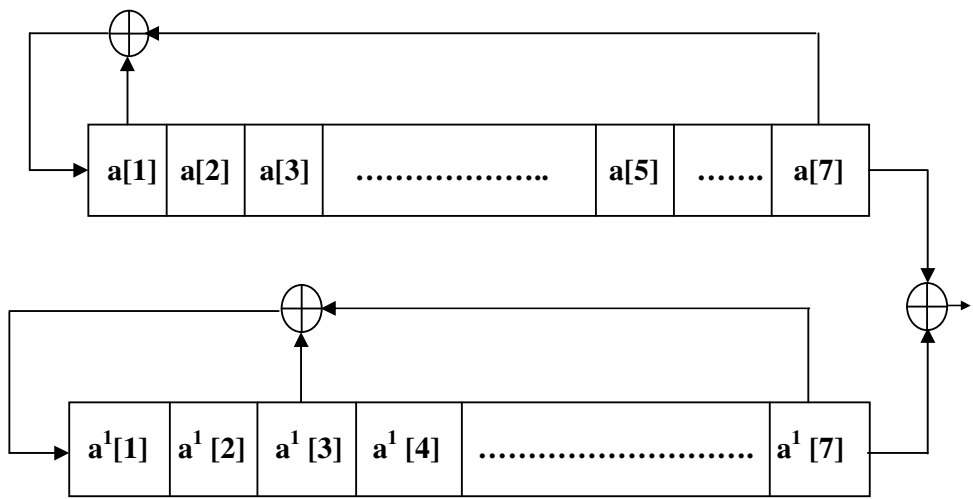

Figure 3. $2^7$-1 Length Gold Code

Here, Gold code of length = 127 is generated using preferred pair of PN codes having characteristic polynomials $X^7+X+1$ and $X^7+X^3+1$ respectively. Thus, the entire gold code family having code length = 7 to 255 were generated through MATLAB programming by simulating the above structure. It has thus been observed that Gold codes have overcome the disadvantage of limited code family size as in PN codes. Also, finding preferred pairs of PN codes is necessary in defining sets of Gold codes.

## 4. EVALUATION OF PEAK CORRELATION CHARACTERISTICS

The two types of Non-orthogonal codes i.e. PN codes and Gold codes are being considered for the evaluation of absolute peak correlation characteristics in this section.

### 4.1. Pseudo-Noise (PN) Codes

The PN codes were generated with length varying from N=7 to 255 using the LFSR structure. Through exhaustive computer simulations, the peak ACF and CCF characteristics of entire PN code family of each length have been evaluated. The ACF characteristics of some of the codes ($X^3+X^2+1$; $X^6+X^1+1$; $X^7+X^1+1$; $X^8+X^6+X^5+X^3+1$) are plotted in Figure 4 for illustration. And the detailed results for each code length are further tabulated in Table 2.

It has been observed from the Figure 4 that all PN codes possess ideal impulsive ACF characteristics. It means that peak value exists at zero time-shift and zero values are present for all other time shifts i.e. no side-lobe exists in ACF characteristics. All possible ACF values obtained through simulations are also tabulated in Table 2. It has also been observed that the peak ACF value is equal to the corresponding PN code length. Thus, PN codes are best in terms of ACF characteristics.

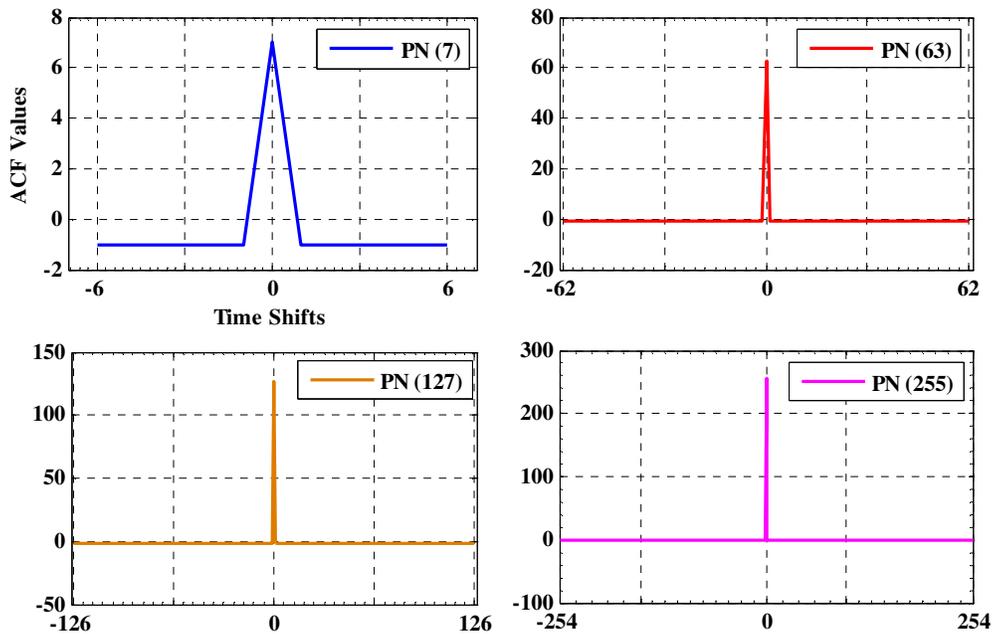

Figure 4. ACF Characteristics of PN Codes of different Lengths

Table 2. ACF & CCF Characteristics of PN Codes

| Length of Code (N) | Code Family Size (M) | ACF Values | CCF Values | Peak Sidelobe ACF wrt N (dB) | Peak CCF wrt N (dB) | % of '-1' CCF values |
|---|---|---|---|---|---|---|
| 7 | 2 | -1, 7 | -5, -1, 3 | No sidelobes (Impulsive ACF) | -7.35 | 46.1 % |
| 15 | 2 | -1, 15 | -5, -1, 3, 7 | No sidelobes (Impulsive ACF) | -6.62 | 34.5 % |
| 31 | 6 | -1, 31 | -9, -1, 7 | No sidelobes (Impulsive ACF) | -12.92 | 49.2 % |
| 63 | 6 | -1, 63 | -17, -1, 15 | No sidelobes (Impulsive ACF) | -12.47 | 74.4 % |
| 127 | 18 | -1, 127 | -17, -1, 15 | No sidelobes (Impulsive ACF) | -18.55 | 49.8 % |
| 255 | 16 | -1, 255 | -33, -17, -1, 15, 31, 63 | No sidelobes (Impulsive ACF) | -12.14 | 40.8 % |

In the similar fashion, the cross-correlation characteristics are also being plotted in Figure 5 and tabulated in Table 2.

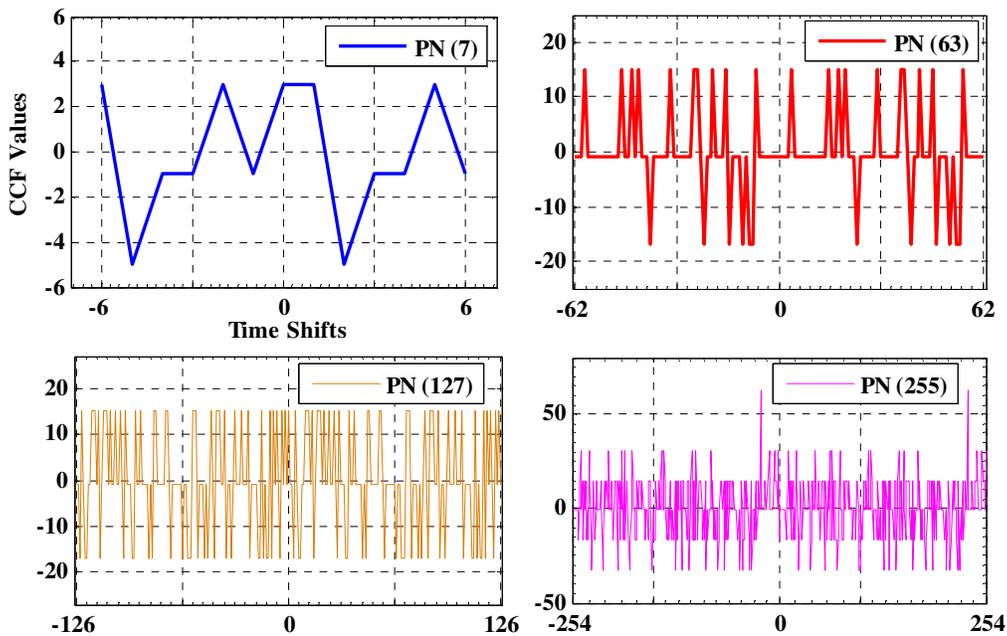

Figure 5. CCF Characteristics of PN Codes of different Lengths

It has been observed from the Figure 5 that none of the PN codes possesses the desirable all zero CCF characteristics. Further, it is observed that the PN sequences generated using preferred pair of feedback polynomials (PN-7, PN-63 and PN-127) yield 3-valued bounded CCF. The bounded CCF values for each length are also being tabulated in Table 2. And it is seen that the peak CCF values for various PN code lengths vary in the range of -6.62 dB to -18.55 dB with respect to ACF peak / Length of the corresponding PN code. Also, this table contains the percentage of '-1' values (equivalent to zero) in CCF for various lengths. The larger the frequency of occurrence of '-1' CCF values, the better is the code. For all lengths of PN codes, the % of '-1' CCF values varies from 34.5% to 74.4%.

### 4.2. Gold Codes

The Gold codes were also generated with length varying from N=7 to 255 by using a pair of PN sequences generated using feedback polynomials listed in section 3.1. The entire set of Gold code family of each length is generated for exhaustive evaluation of peak ACF and CCF characteristics. The auto-correlation and cross-correlation characteristics have been evaluated for every Gold code through MATLAB programming and the results are tabulated in Table 3.

Table 3.  ACF & CCF Characteristics of Gold Codes

| Length of Code (N) | Code Family Size (M) | ACF Values | CCF Values | Peak Sidelobe ACF wrt N (dB) | Peak CCF wrt N (dB) | % of '-1' CCF values |
|---|---|---|---|---|---|---|
| 7 | 9 | -5, -1, 3, 7 | -5, -1, 3 | -7.35 | -7.35 | 53 % |
| 15 | 17 | -5, -1, 3, 7, 15 | -5, -1, 3, 7 | -6.62 | -6.62 | 39.4 % |
| 31 | 33 | -9, -1, 7, 31 | -9, -1, 7 | -12.92 | -12.92 | 50.8 % |
| 63 | 65 | -17, -1, 15, 63 | -17, -1, 15 | -12.47 | -12.47 | 75.2 % |
| 127 | 129 | -17, -1, 15, 127 | -17, -1, 15 | -18.55 | -18.55 | 50.2 % |
| 255 | 257 | -33, -17, -1, 15, 31, 63, 255 | -33, -17, -1, 15, 31, 63 | -12.14 | -12.14 | 41.2 % |

It has already been mentioned that Gold code family of each length also includes the pair of PN codes used for their generation. These two numbers of Gold codes therefore inherits the same correlation characteristics of PN codes. The remaining Gold codes however, possess different correlation characteristics. In terms of ACF, it is seen from the Table 3 that none of these Gold codes possess the ideal 2-valued impulsive characteristics. The ACF sidelobes also assume significance in case of Gold codes and the level of these sidelobes vary from -6.62 dB to -18.55 dB with respect to ACF peak / Length of corresponding code.

In order to bring out a clear picture of ACF characteristics of Gold codes, ACF of all possible Gold codes of length 7 are plotted in Figure 6. It has been observed from Figure 6(b) – (h) that no Gold code has two valued impulsive characteristics. The sidelobes with respect to main peak at the centre are clearly noticed in each graph. Figure 6(a) however illustrates the impulsive ACF of generating PN sequences which are a part of Gold code family.

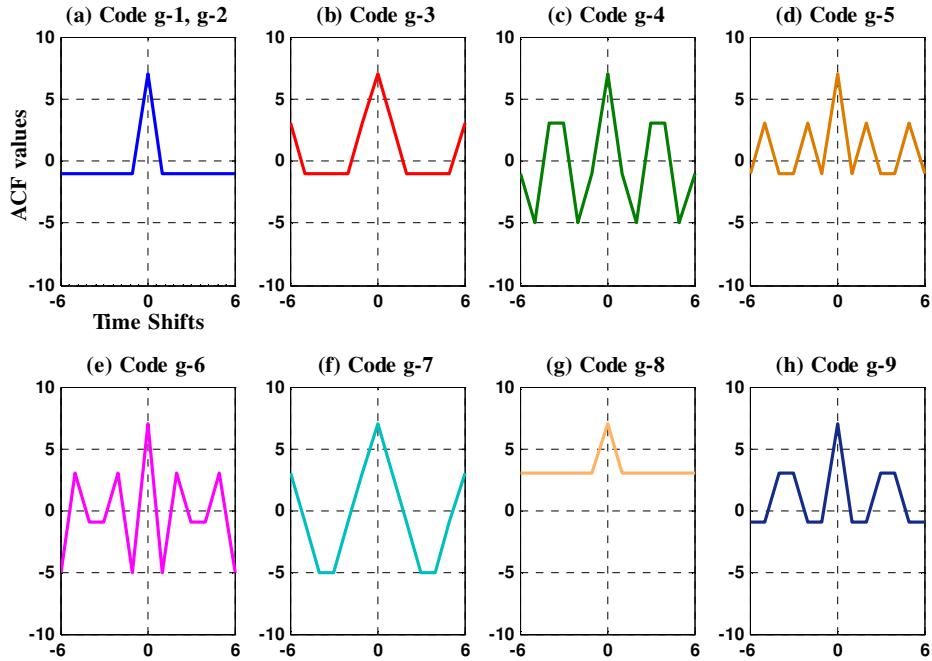

Figure 6. ACF Characteristics of all Gold Codes of Length 7

Further, it is clearly seen from Table 3 that peak CCF sidelobes' level in case of Gold codes are exactly same as in Table 2 of PN codes. Other observation that can be made with regard to CCF characteristics is that Gold codes possess bounded CCF values and there exists pairs of Gold sequences which guarantee better CCF characteristics than their PN counterpart. It is also observed from Table 3 that for all lengths of Gold codes, the % of '-1' CCF values varies from 39.4% to 75.2%.

## 5. COMPARATIVE EVALUATION OF PEAK CORRELATION CHARACTERISTICS

The detailed evaluation of the PN codes and Gold codes with regard to peak ACF & CCF characteristics has already been done in the previous section. In terms of peak ACF characteristics, PN codes are best and provide desirable 2-valued impulsive characteristics. Gold codes, on the other hand do not have impulsive ACF. The ACF sidelobes assume significance in case of Gold codes and the level of these sidelobes vary from -6.62 dB to -18.55 dB wrt to ACF main peak. Further, a look at the evaluated CCF characteristics of the two codes reveals that their CCF characteristics are more or less identical. It is clearly seen from Table 3 that peak CCF sidelobes' level in case of Gold codes are exactly same as in Table 2 of PN codes.

One significant observation that can be made here is that Gold codes possess bounded CCF values and there exists pairs of Gold sequences which guarantee better CCF characteristics than their PN counterpart. This fact is also supported by the frequency of occurrence of '-1' CCF values in Gold codes of various lengths. It is clearly observed from Table 3 that for all lengths, Gold codes have larger % of '-1' CCF values as compared to PN codes (Table 2). Therefore, it is concluded that Gold codes are better than PN codes in terms of peak CCF characteristics. Apart from CCF characteristics, the family size of Gold codes (Table 3) is also significantly larger than PN codes (Table 2). Thus, it is concluded that except ACF, Gold codes offer many advantages over their PN counterpart. Though Gold codes do not possess impulsive ACF characteristics, but the ACF sidelobes of Gold codes are significantly lower than main ACF peak. The level of these sidelobes varies from -6.62 dB to -18.55 dB wrt to the ACF main peak.

In order to highlight clearly the difference in ACF and CCF characteristics of PN and Gold codes, the graphs are plotted in Figures 7 and 8 respectively for code length = 256.

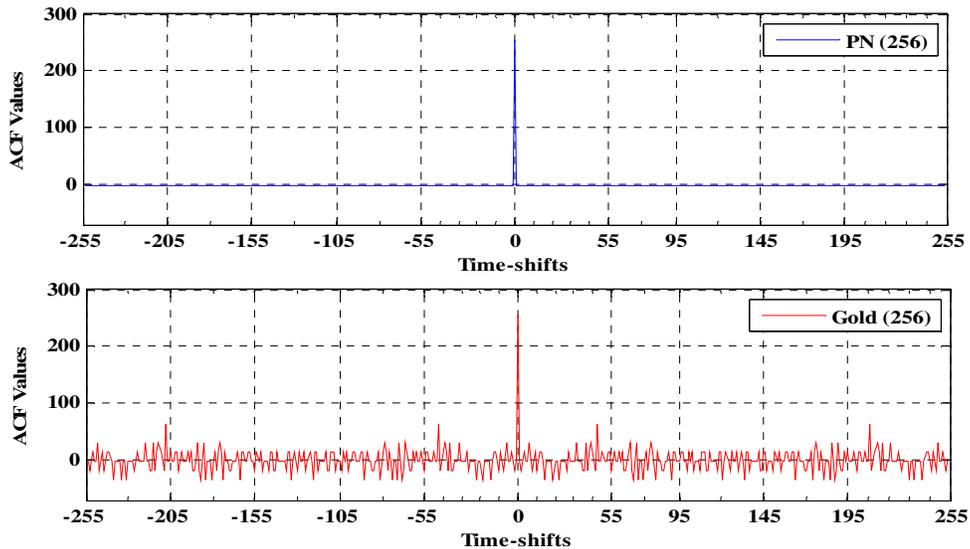

Figure 7. Comparative ACF Characteristics of PN & Gold Codes of Length 256

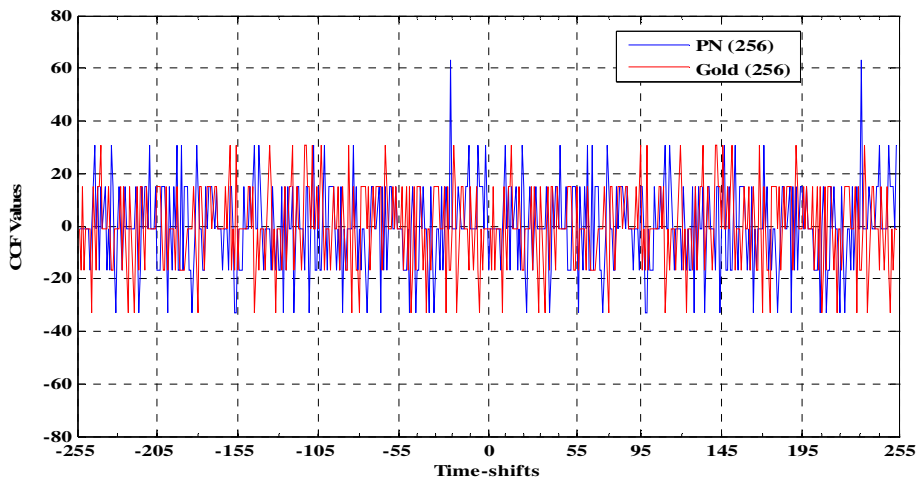

Figure 8. Comparative CCF Characteristics of PN & Gold Codes of Length 256

Figure 7 clearly shows that the PN codes possess impulsive ACF characteristics. On the other hand, Gold codes have multi-valued ACF but peak sidelobe ACF is 12.14 dB down with respect to main peak of ACF. Further, it is observed from Figure 8 that the CCF values of Gold codes are bounded and their maximum, positive peak CCF values are also 6 dB down wrt maximum peak CCF of PN codes. Thus, overall among non-orthogonal category, Gold codes are a better candidate for application in CDMA based next generation wireless system.

## 6. CONCLUSIONS

While designing a CDMA system, the proper choice of spreading codes is of prime importance because the performance of a CDMA based wireless system is mainly governed by the characteristics of spreading sequences. Even for hybrid CDMA systems, suppressing the Multiple Access Interference (MAI) is a crucial problem. The imperfect correlation characteristics of spreading codes and the multipath fading destroy the orthogonality among various users and thus the resulting MAI produces serious BER degradation in the system. The desirable characteristics of CDMA spreading codes include (i) availability of large number of codes (ii) impulsive auto-correlation function (iii) zero cross-correlation value (iv) low Peak to Average Power Ratio (PAPR) value and (v) support for variable data rates.

Ideally, a spreading code should possess both impulsive auto-correlation (zero sidelobe levels) and all zero cross-correlation characteristics. But unfortunately, no such code family exists which possess both correlation characteristics simultaneously. Therefore, in order to have tolerable MAI, one has to compromise with maximum possible difference between ACF (auto-correlation function) peak and CCF (cross-correlation function) peak of the codes selected. Keeping this in view, in this paper various Non-orthogonal codes (PN code and Gold code) are investigated with special emphasis on their peak correlation characteristics.

Exhaustive simulation has proved that in terms of peak ACF characteristics, PN codes are best and provide desirable 2-valued impulsive characteristics. Gold codes, on the other hand do not have impulsive ACF. The ACF sidelobes assume significance in case of Gold codes and the level of these sidelobes vary from -6.62 dB to -18.55 dB wrt to ACF main peak. However, the peak ACF sidelobe level of Gold codes, though not zero, is quite tolerable. Further, Gold codes possess bounded CCF values and there exists pairs of Gold sequences which guarantee better CCF characteristics than their PN counterpart. The positive peak CCF values for Gold codes are significantly down (approx. 6 dB) wrt maximum peak CCF of PN codes. Further, for all lengths, Gold codes have larger % of '-1' CCF values (39.4% to 75.2%) as compared to PN codes (34.5% to 74.4%). Therefore, it is concluded that Gold codes are better than PN codes in terms of peak CCF characteristics. Apart from CCF characteristics, the family size of Gold codes is also significantly larger than PN codes (Table 2).Therefore, it is concluded that overall among non-orthogonal category, Gold codes are a better candidate for application in CDMA based wireless system.

**Author**

Deepak Kedia received his B.E. degree with Hons. in Electronics & Communication Engineering in 2000 from DCRUST, Murthal (India), M.Tech (Telecommunication) in 2002 from IIT, Kharagpur and Ph.D in the field of Wireless mobile communication from GJUS&T, Hisar in 2011. Currently, he is working as a Senior Assistant Professor in ECE Department at GJUS&T, Hisar (India). He has an experience of 10 years in teaching and research. His research interests include mobile communication and spread spectrum.

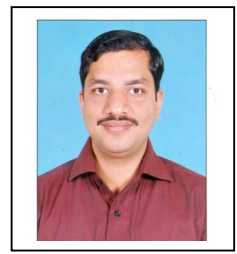